\documentclass[aps,preprintnumbers,amsmath,amssymb,superscriptaddress,floatfix,nofootinbib]{revtex4}
\usepackage{lipsum}
\usepackage{cancel}
\usepackage{slashed}
\usepackage{graphicx}
\usepackage{epsfig}
\usepackage{comment}
\usepackage{epstopdf}
\usepackage{hyperref}
\usepackage{amsmath}
\usepackage{amsfonts}
\usepackage{amssymb}
\usepackage{setspace}
\usepackage{amsfonts,color}
\usepackage{caption}
\usepackage{natbib}
\usepackage{subfig}
\usepackage{color}
\usepackage{bbold}
\usepackage{bm}
\usepackage{mathrsfs}
\usepackage{url}
\usepackage{rotating}
\usepackage{pdflscape}
\usepackage{diagbox}
\usepackage[normalem]{ulem}
\newcommand{\red}[1]{\textcolor{red}{#1}}

\usepackage{multirow}
\usepackage{hhline}
\usepackage{makecell}
\usepackage{float}

\begin{document}

\title{Kaluza-Klein inspired a model of the inflation with the inversed power law potential in Bianchi type-I universe}

\author{Jaturaporn Wattanakumpolkij}
\email{jaturaporn.wa@kkumail.com}
\affiliation{Khon Kaen Particle Physics and Cosmology Theory Group (KKPaCT), Department of Physics, Faculty of Science, Khon Kaen University,123 Mitraphap Rd., Khon Kaen, 40002, Thailand}
\author{Patinya Ma-ardlerd}
\email{m.patinya@kkumail.com}
\affiliation{Khon Kaen Particle Physics and Cosmology Theory Group (KKPaCT), Department of Physics, Faculty of Science, Khon Kaen University,123 Mitraphap Rd., Khon Kaen, 40002, Thailand}
\author{Natthason Autthisin}%
\email{natthasorn\_ut@kkumail.com}
\affiliation{Khon Kaen Particle Physics and Cosmology Theory Group (KKPaCT), Department of Physics, Faculty of Science, Khon Kaen University,123 Mitraphap Rd., Khon Kaen, 40002, Thailand}
\author{Pornpatara Chuvala}
\email{pornpatara.ch@mail.wu.ac.th}
\affiliation{School of Science, Walailak University, Thasala, Nakhon Si Thammarat, 80160, Thailand}
\affiliation{Khon Kaen Particle Physics and Cosmology Theory Group (KKPaCT), Department of Physics, Faculty of Science, Khon Kaen University,123 Mitraphap Rd., Khon Kaen, 40002, Thailand}
\author{Nutthaphat Lunrasri}
\email{natthapatl@kkumail.com}
\affiliation{Khon Kaen Particle Physics and Cosmology Theory Group (KKPaCT), Department of Physics, Faculty of Science, Khon Kaen University,123 Mitraphap Rd., Khon Kaen, 40002, Thailand}
\author{Chakrit Pongkitivanichkul}
\email{chakpo@kku.ac.th}
\affiliation{Khon Kaen Particle Physics and Cosmology Theory Group (KKPaCT), Department of Physics, Faculty of Science, Khon Kaen University,123 Mitraphap Rd., Khon Kaen, 40002, Thailand}
\author{Daris Samart}
\email{darisa@kku.ac.th, corresponding author}
\affiliation{Khon Kaen Particle Physics and Cosmology Theory Group (KKPaCT), Department of Physics, Faculty of Science, Khon Kaen University,123 Mitraphap Rd., Khon Kaen, 40002, Thailand}

\date{\today}

\begin{abstract}
This work considers the dynamics of the gauge vector and inflaton (dilaton) fields inspired by Kaluza-Klein theory in an inflationary universe with Bianchi type-I spacetime. The inverse power-law potential of the inflaton field is used to study dynamical system analysis. As a result, all fixed points in the autonomous system are non-hyperbolic fixed points, and one cannot determine their stability. Therefore, a center manifold theory is required to analyze the stability of the dynamical system properly. Interestingly, we found an isotropic attractor point which means that the universe undergoes accelerated expansion (inflation) from an anisotropic phase to an isotropic phase of the universe. According to the dynamical system analysis of the anisotropic Bianchi type-I universe with the inspired Kaluza-Klein model, our results supported the isotropization of the observed universe.
\end{abstract}

\maketitle
\section{Introduction}\label{sec-1}

The study of the early universe has long attempted to address fundamental questions regarding its initial conditions and subsequent evolution. A cornerstone in this effort is the theory of cosmic inflation, which postulates a period of quasi-exponential expansion that resolves several conundrums of the standard Big Bang model \cite{Clifton:2011jh}. While the standard inflationary framework, constructed within the isotropic and homogeneous Friedmann-Lemaître-Robertson-Walker (FLRW) spacetime, successfully explains the large-scale uniformity of the cosmic microwave background (CMB) and the origin of cosmic structure, it is crucial to test its robustness by considering more general initial conditions. Extending inflationary models to include anisotropic geometries allows us to investigate the cosmic no-hair conjecture, which posits that inflation naturally smooths out and isotropizes the universe, regardless of its initial state \cite{Gron:1985hw, Lim:2003ut, Pereira:2015pga}. The study of such extensions is vital to understand how our universe, which may have emerged from highly anisotropic beginnings, achieved its present state of remarkable isotropy.

While the cosmic no-hair conjecture holds in many standard scenarios, it can be evaded in certain modified gravity theories or models involving non-trivial field dynamics \cite{Clifton:2011jh, Nojiri:2022idp}. In particular, models of anisotropic inflation driven by vector fields have demonstrated that a small, persistent anisotropy can be maintained throughout the inflationary epoch. Seminal works showed that vector fields, whether through non-minimal couplings or inherent "vector impurity," could source a stable anisotropic expansion \cite{PhysRevD.40.967, Golovnev:2008cf, Kanno:2008gn, Watanabe:2009ct}. This concept was further developed into various models, including power-law, constant-roll, and hyperbolic anisotropic inflation \cite{Kanno:2010nr, Soda:2012zm, Ito:2017bnn, Chen:2021nkf}. The dynamics of these scenarios have been explored extensively, considering different types of fields such as non-Abelian gauge fields, form fields, and models with non-minimal kinetic coupling or warm inflationary effects \cite{Murata:2011wv, Maleknejad:2011jw, Maleknejad:2012fw, Ito:2015sxj, Goodarzi:2022wli, Kanno:2022flo, Pham:2023evo}. These investigations have revealed that anisotropic inflationary solutions can be sustained, potentially leaving detectable signatures such as statistical anisotropy in the CMB \cite{Pitrou:2008gk, Soda:2012zm, Soda:2014awa} and influencing the production of primordial gravitational waves \cite{Soda:2012zm, Chen:2021nkf}. A significant body of recent work continues to refine these models, exploring scenarios with multiple scalar and vector fields and their embedding within alternative theories of gravity \cite{Do:2021lyf, Do:2021pqk, Do:2023mqe, Do:2025xhm}.

To investigate these dynamics, Bianchi type cosmologies, which preserve homogeneity but allow for directional scale factors, provide the natural theoretical laboratory. The Bianchi type-I model is the simplest such framework, and it has been the setting for much of the foundational research into anisotropic inflation. A key requirement for any such model is the existence of a viable isotropization mechanism, ensuring that the anisotropic phase gracefully transitions into the isotropic universe we observe today. Anisotropic models, while allowing for small anisotropies during inflation, must admit attractor solutions that drive the universe toward isotropy at late times. The conditions under which this transition occurs, especially in the presence of complex matter contents, demand a rigorous and systematic analysis.

On a parallel front, theories extending general relativity, such as Kaluza-Klein (KK) theory, have long been pursued to unify gravity with other fundamental forces \cite{Overduin:1997sri, Bailin:1987jd}. KK theory posits the existence of extra spatial dimensions compactified to unobservable scales. Early explorations of KK cosmology already laid the groundwork for understanding its cosmological implications \cite{Freund:1982pg, Randjbar-Daemi:1983awk, Osorio:1993iv}. In the modern context, the dimensional reduction of a higher-dimensional metric naturally yields a four-dimensional effective theory containing not only gravity but also a scalar field (the dilaton) and one or more gauge vector fields. These emergent fields provide a compelling physical motivation for the ingredients often used in anisotropic inflation models. Indeed, recent KK-inspired Brans-Dicke and $f(R,T)$ models have been proposed to explain dark matter, dark energy, and the universe's accelerated expansion, highlighting the continued relevance of the KK paradigm \cite{Sahoo:2014wgz, Pongkitivanichkul:2020txi, SurendraSingh:2021uxb, Waeming:2021ytf, Jusufi:2024utf}.

To analyze the complex, nonlinear behavior of cosmological models with multiple fields and anisotropies, dynamical systems theory has become an indispensable tool. This approach reformulates the cosmological field equations into an autonomous system of first-order differential equations, allowing for a qualitative understanding of the universe's complete evolution through the analysis of fixed points and phase space trajectories. The power of this method has been demonstrated in numerous comprehensive reviews and studies covering dark energy, modified gravity, and various scalar field potentials \cite{Garcia-Salcedo:2015ora, Tamanini:2014mpa, Alho:2015cza, Bouhmadi-Lopez:2016dzw, Boehmer:2014vea, Bahamonde:2017ize, Pal:2022jak, Alho:2022ehq, Clarkson:2001fd}. Specifically, it has been used to study the dynamics of exponential potentials for late-time cosmology \cite{Copeland:1997et, Heard:2002dr, Guo:2003eu, Clarkson:2001fd} as well as steep potentials relevant for early-time inflation \cite{Das:2019ixt}, and also leads to power-law inflation unaffected by magnetic fields \cite{Clarkson:2001fd}. This framework has also been successfully applied to analyze anisotropic cosmological models, clarifying the stability of their isotropic and anisotropic phases \cite{Chaubey:2016qtx, Karciauskas:2016pxn, Ghaffarnejad:2017hqt, Guarnizo:2020pkj, Bhanja:2023jro, Do:2023yvg}.

A critical limitation of standard dynamical systems analysis arises when the fixed points of the system are non-hyperbolic—that is, when the Jacobian matrix evaluated at the fixed point has one or more eigenvalues with a zero real part. In these cases, linear stability analysis is inconclusive. Center Manifold Theory (CMT) provides a rigorous mathematical technique to overcome this issue by reducing the dynamics to a lower-dimensional, stable manifold (the ``center manifold") on which the long-term behavior of the system can be definitively determined \cite{Rendall:2001it, Alho:2014fha}. CMT has proven crucial for establishing the stability of cosmological solutions in a variety of contexts where non-hyperbolic points appear, such as in models with non-minimal couplings \cite{Hrycyna:2015eta}, scalar field attractors \cite{Alho:2015cza}, and phantom crossing scenarios \cite{Boehmer:2011tp, Pal:2019qch}. These studies underscore the necessity of CMT for a complete and accurate stability analysis when standard methods fail.

In this work, we construct a novel inflationary model inspired by Kaluza-Klein theory within a Bianchi type-I spacetime. The model includes a scalar inflaton (dilaton) field governed by an inverse power-law potential and a vector field, both arising naturally from the KK dimensional reduction. We reformulate the field equations as an autonomous dynamical system and find that all of its critical points are non-hyperbolic. Consequently, we employ Center Manifold Theory to rigorously analyze their stability. Our main result is the identification of a stable isotropic attractor solution, which demonstrates that the universe can dynamically evolve from an anisotropic state to an isotropic inflationary phase, providing a concrete mechanism for isotropization. To the best of our knowledge, this work represents the first analysis of anisotropic inflation with an inverse power-law potential that uses Center Manifold Theory to assess the stability of non-hyperbolic fixed points. This combination of anisotropic cosmology, KK-inspired vector-scalar fields, and advanced dynamical system analysis fills an important gap in the literature, contributing new insights into the isotropization mechanism in early universe cosmology.

This paper is organized as follows. In Sec.~II, we introduce the KK-inspired action and derive the field equations in the Bianchi type-I spacetime. In Sec.~III, we formulate the autonomous dynamical system and analyze its fixed points. In Sec.~IV, we employ Center Manifold Theory to determine the stability of the non-hyperbolic fixed points. Finally, in Sec.~V, we summarize our results and discuss their implications for early universe cosmology.

\newpage
\section{KK inspired inflation model in Bianchi type-I universe}
This section aims to formulate the dynamical equations of the model. We will start with the KK inspired model action, and the cosmological equations in the Bainchi type-I. The KK theory is a framework that unifies gravity with electromagnetism by introducing an extra spatial dimension. The idea is that the metric tensor in 5 dimensions (5D) can be decomposed into a 4-dimensional (4D) metric, a vector field, which corresponds to the electromagnetic potential, and a scalar field, sometimes called the dilaton. The original KK action in 5D reduces to a 4D action containing gravity, a Maxwell field, and a scalar field. Then the action is given by,
\begin{eqnarray}
  S = \int{}d^4\text{x}\sqrt{-g}\bigg[{1\over2}M_p^2\,R-{1\over2}g^{\mu\nu}\partial_\mu\phi\partial_\nu\phi-{1\over4}f(\phi)F_{\mu\nu}F^{\mu\nu}-V(\phi)\bigg]
\end{eqnarray} 
Where $M_p=1/\sqrt{8\pi G}$ is a reduced Planck mass and $F_{\mu\nu}$ is the field strength of the vector field defined by $F_{\mu\nu}=\partial_\mu A_\nu - \partial_\nu A_\mu$ with a coupling function, $f(\phi)=e^{-\sqrt{3}\phi/M_p}$ where $A_\mu$ represents the vector field whose kinetic term couples with inflation. One notes that the factor $\sqrt{3}$ form the coupling function $f(\phi)$ has its own uniqueness due to the dimensional reduction from 5D KK gravity to 4D. The anisotropic Universe allowed us to give the field propagating in the $x$-exist $A_\mu = (0,A_x(t),0,0)$ and holds the plane symmetry in the plane perpendicular to the propagating direction. Here, we consider the Bianchi type-I spacetime model having line elements as follows: 
\begin{eqnarray}
  ds^2 = -dt^2+e^{2\alpha(t)}\Big[e^{-4\sigma(t)}dx^2+e^{2\sigma(t)}(dy^2+dz^2)\Big]
\end{eqnarray}
Where $t$ is the cosmic time. The isotropic scale factor $a(t)$ is defined as $e^{\alpha(t)}$ and the shear tensor $\sigma(t)$ represents a deviation from the isotropy. We can obtain the basic equations by taking variations of the action,
\begin{eqnarray}
\dot{\alpha}^2(t)&=&
\dot{\sigma}^2(t)+{1\over3M_p^2}\bigg[{1\over2}e^{-2\alpha(t)+4\sigma(t)}f^{2}(\phi)\dot{A_x}^2(t)+{1\over2}\dot{\phi}^2(t)+V(\phi)\bigg]
\label{eq:FriedmannEq}\\
\ddot{\alpha}(t)&=&
-3\dot{\alpha}^2(t)+{1\over6M_p^2}e^{-2\alpha(t)+4\sigma(t)}f^{2}(\phi)\dot{A_x}^2(t)+{V(\phi)\over M_p^2}\\
\ddot{\sigma}(t)&=&
-3\dot{\alpha}(t)\dot{\sigma}(t)+{1\over3M_p^2}e^{-2\alpha(t)+4\sigma(t)}f^{2}(\phi)\dot{A_x}^2(t)\\
\ddot{\phi}(t)&=& -3\dot{\phi}(t)\dot{\alpha}(t)-V'(\phi)+e^{-2\alpha(t)+4\sigma(t)}f(\phi)f'(\phi)\dot{A_x}^2(t)
\end{eqnarray}
Where $\dot{A_x}(t)=f^{-2}(\phi)e^{-\alpha(t)-4\sigma(t)}p_A$, $p_A$ is the integral constant.

\section{Dynamical system analysis and center manifold theory}
To analyze the entire evolution of the system, we define a set of dimensionless variables normalized by the mean expansion rate ($H=\dot{\alpha}$),
\begin{eqnarray}
S={\dot{\sigma}(t)\over\dot{\alpha}(t)},~~X={\dot{\phi}(t)\over\sqrt{6}M_p\dot{\alpha}(t)},~~Y={\sqrt{V(\phi)}\over\sqrt{3}M_p\dot{\alpha}(t)},~~\lambda=-{M_pV'(\phi)\over V(\phi)}
\end{eqnarray}
In terms of these variables, the Friedmann constraint takes the compact form,
\begin{eqnarray}
    1 = S^2 + X^2 + Y^2 + \Omega_A,
    \label{eq:FriedmannCon}
\end{eqnarray}
where the density parameter for the vector field is defined as
\begin{eqnarray}
    \Omega_A = \frac{1}{6M_p^2\dot{\alpha}^2(t)}\,e^{-4\alpha(t)-4\sigma(t)}f^{-2}(\phi) p_A^2,
\end{eqnarray}
and $\Omega_\sigma=S^2$ is the density parameter for shear (anisotropy), $\Omega_\phi=X^2+Y^2$ is scalar field density parameter. The energy density and pressure for the scalar field are given by
\begin{eqnarray}
    \rho_\phi &=& {1\over2}\dot{\phi}(t)^2 + V(\phi), \\
    P_\phi &=& {1\over2}\dot{\phi}(t)^2 - V(\phi).
\end{eqnarray}
Using these definitions, the equation of state of the scalar field can then be written as
\begin{eqnarray}
    w_\phi = {P_\phi \over \rho_\phi} = \frac{X^2 - Y^2}{X^2 + Y^2}.
\end{eqnarray}
With the density parameters for both the vector field and the scalar field established, we can define the effective equation of state (EoS) parameter of the universe as the ratio of the total pressure to the total energy density
\begin{eqnarray}
    w_{\rm eff} = \frac{P_{\rm eff}}{\rho_{\rm eff}} = \frac{P_\phi + P_{\sigma} + P_A}{\rho_\phi + \rho_{\sigma} + \rho_A} = X^2-Y^2+w_{\sigma}S^2+w_A(1-S^2-X^2-Y^2),
\end{eqnarray}
Where $w_A$ is the equation of state of the vector field and $w_\sigma$ represents the equation of state of anisotropic expansion. The deceleration parameter measures the rate of change of the expansion rate
\begin{eqnarray}
    q = - \left( 1 + \frac{\ddot{\alpha}(t)}{\dot{\alpha}(t)^2} \right) = 1+S^2+X^2-2Y^2.
\end{eqnarray}
Using e-folding number as a time coordinate $d\alpha=\dot{\alpha}dt$, the basic equations can be reduced to the autonomous form:
\begin{eqnarray}
{dS\over d\alpha} &=& 
-3SY^2+(2-S)(1-S^2-X^2-Y^2),  \\
{dX\over d\alpha} &=&
-3XY^2+\sqrt{{3\over2}}Y^2\lambda-(X+3\sqrt{2})(1-S^2-X^2-Y^2),\\
{dY\over d\alpha} &=& 
3Y-3Y^3-\sqrt{{3\over2}}XY\lambda-Y(1-S^2-X^2-Y^2),\\
{d\lambda\over d\alpha} &=& 
\sqrt{6}X\lambda^2-\sqrt{6}X\Gamma\lambda^2,
\end{eqnarray}
where $\Gamma\equiv {VV_{\phi\phi}/V_\phi^2}$ which $V_{\phi}={dV\over d\phi}$ and $V_{\phi\phi}={d^2V\over d\phi^2}$. The Friedmann constraint enabled us to compactify the phase space. Note that $-1\leq X\leq 1, 0\leq Y\leq 1 ~~\text{and}~~ 0\leq S\leq 1$ but the $\lambda$ is not in the constraint. To make the $\lambda$ is compact, we can motivated by define $Z=\lambda/(\lambda+1)$. Consequently $0\leq Z\leq 1$, and we need not worry about the presence of critical points at infinity, thus, we multiply the right-hand side of the autonomous equations by $(1-Z)$, the autonomous system becomes
\begin{eqnarray}\label{autonomous sysytem}
S' &=& 
-\Big((S-2)(S^2+X^2-1)-2(1+S)Y^2\Big)(Z-1), \label{autonomous sysytem1}\\
X' &=&
(1-Z)\left(-3XY^2-(X+3\sqrt{2})(1-S^2-X^2-Y^2)\right)+\sqrt{3\over2} Y^2Z ,\label{autonomous sysytem2}\\
Y' &=& 
{1\over2}Y(Z-1)\Big(-2S^2-2X^2+4(Y^2-1)\Big)-{\sqrt{6}\over2}XYZ, \label{autonomous sysytem3}\\
Z' &=& {\sqrt{6}\over n}\,X(Z-1)Z^2.
\label{autonomous sysytem4}
\end{eqnarray}
Here, a prime denotes derivative with respect to $\alpha$. Note that the power-law potential $V(\phi)=M^{n+4}\phi^{-n}$ yields the parameter $\Gamma=(n+1)/n$.
To determine the dynamics of the system, it is need to find out the fixed points, which are obtained from the simultaneous solutions of the equations $S'=0, X'=0, Y'=0$ and $Z'=0$. The stability of a fixed point can be determined by examining the eigenvalues of the Jacobian matrix, which is composed of the first-order partial derivatives of the autonomous system and is evaluated at the fixed points. The fixed point of our system, along with its eigenvalues, is shown in Table \ref{table:1}. The fixed points along with the cosmological parameters are displayed in Table \ref{table:cosmoparameters}.
\begin{table}[h!]
\setlength{\tabcolsep}{7pt}
\renewcommand{\arraystretch}{1.7}
\begin{tabular}{| c | c | c | c | c | c | c | c |}
\hline
Point & S & X & Y & Z & Existence & $\Omega_A$ & Eigenvalues\\
\hline
A & Any & Any & 0 & 1 & $S^2+X^2\leq 1$ & $1-S^2-X^2$ & $\Big\{0,0,-{\sqrt{3\over2}}X,{\sqrt{6}\over n}X\Big\}$ \\

B & Any & 0 & 0 & 1 & $S
^2\leq 1$& $1-S^2$ & $\{0,0,0,0\}$\\

C & 1 & 0 & 0 & Any & $0\leq Z \leq 1$ & 0 & $\{2(Z-1),0,-3(Z-1),0\}$\\

D & $\sqrt{1-X^2}$ & Any & 0 & 0 &$-1\leq X \leq 1$ & 0 & $\{0,2+6\sqrt{2}X-4\sqrt{1-X^2},3,0\}$\\

E & 0 & 0 & 1 & 0 & All values & 0 & $\{-3,-3,-4,0\}$\\
\hline
\end{tabular}
\caption{The fixed points and their corresponding eigenvalues in this system.}
\label{table:1}
\end{table}

\begin{table}[h!]
\setlength{\tabcolsep}{7pt}
\renewcommand{\arraystretch}{1.7}
\begin{tabular}{| c | c | c | c | c | c | c | c |}
\hline
Point & S & X & Y & Z & $\Omega_A$ & $w_{\rm eff}$ & $q$ \\
\hline
A & Any & Any & 0 & 1 & $1-S^2-X^2$ & $w_{\sigma}S^2+X^2+w_A(1-S^2-X^2)$ &  $1+S^2+X^2$\\

B & Any & 0 & 0 & 1 & $1-S^2$ & $w_{\sigma}S^2+w_A(1-S^2)$ & $1+S^2$ \\

C & 1 & 0 & 0 & Any & 0 & $w_\sigma$ & 2 \\

D & $\sqrt{1-X^2}$ & Any & 0 & 0 & 0 & $X^2+w_{\sigma}(1-X^2)$ & 2 \\

E & 0 & 0 & 1 & 0 & 0 & $-1$ & $-1$ \\
\hline
\end{tabular}
\caption{The fixed points and cosmological parameters for our system.}
\label{table:cosmoparameters}
\end{table}

For a hyperbolic point, stability around the fixed point depends on the sign of the eigenvalues, provided all eigenvalues have a non-zero real part. In our case, all the fixed points have at least one zero eigenvalue, indicating that the points are non-hyperbolic, and linear stability analysis is not applicable. To analyze the stability of the non-hyperbolic points, we can use the Center Manifold Theorem (CMT).

We begin by reviewing a mathematical description of the basics of CMT.The main concept of CMT involves reducing the dimension of the system near the fixed point, enabling an examination of the stability of the reduced system. For $x\in \mathbb{R}^c$ and $y \in \mathbb{R}^s$, an arbitrary dynamical system with zero eigenvalues in the stability matrix can always be written in the following form:
\begin{eqnarray}\label{a1}
x'=Ax+f(x,y)\\
y'=By+g(x,y)
\end{eqnarray}
Where $A$ is a square matrix contained eigenvalues with zero real part and $B$ is a square matrix having eigenvalues with negative real parts. Here we assume that fixed point located at the origin $(0,0)$ where the first derivatives of the vector function $f$ and $g$ satisfy
\begin{eqnarray}
f(0,0)=0,~~Df(0,0)=0\\
g(0,0)=0,~~Dg(0,0)=0
\end{eqnarray}
By definition: if the center manifold is locally it can be represented by $W^c(0)= \lbrace(x,y)\in\mathbb{R}^c\times\mathbb{R}^s|y=h(x),|x|<\delta,h(0)=0,Dh(0)=0 \rbrace$ for $\delta$ sufficiently small. From the definition imply that the space $W^c(0)$ is tangent to the eigenspace at the critical point $(x, y)=(0, 0)$. Thus, the quasilinear partial differential equation is following:
\begin{eqnarray}\label{tangencycond}
\mathcal{N}(h(x))\equiv Dh(x)\left[Au+f(x,h(x))\right]-Bh(x)-g(x,h(x))=0
\end{eqnarray}
In general, we cannot find a solution to this equation. Instead, we can only approximate the function $h(x)$ through a series expansion by ignoring the linear term and constant term, as following the form:
\begin{eqnarray}
h(x) = ax^2+bx^3+\mathcal{O}(x^4)
\end{eqnarray}
Then substituting the approximated solution of $h(x)$ into (\ref{tangencycond}) and comparing the coefficient of  each power of $x$. The dynamics of the center manifold is given by
\begin{eqnarray}
x'= Ax+f(x,h(x))
\end{eqnarray}
Consider the lowest order in the final expression can be written as the form $u'=ku^n$ where $k$ is a constant. If $k<0$ and $n$ is an odd number, it indicates that the original system is stable. Other cases indicate system instability.

\subsection{Point A (Any, Any, 0, 1)}

All the points with $Y=0$ and $Z=0$ are fixed points for any value of $S$ and $X$. This implies that the $S$-axis and $X$-axis are fixed lines in phase space in the range $0 \leq S \leq 1$ and $-1 \leq X \leq 1$. The scalar field has no contribution from its potential energy $(Y=0)$, meaning the scalar field is entirely kinetic. The steepness parameter $Z=1$ corresponds to $\lambda=\infty$. The density parameter, the deceleration parameter, and the effective EoS depend on the $S$ and $X$ representing that anisotropy and kinetic energy may play a role. Since the deceleration parameter $q>0$ for all values of $S$ and $X$, this point always represents a decelerating Universe. For an isotropic case $(S=0)$, if the kinetic energy of the scalar field is dominant $(X\not=0)$, then $\Omega_\phi=1$, meaning the scalar field behaves like a stiff matter $(w_{\rm eff}=1)$. If kinetic energy plays no role $(X=0)$, the energy density is dominated by the vector field, the EoS $w_{\rm eff}=w_A=1/3$ which is characteristic of the equation of state that is similar to radiation. For an anisotropic case, $(S=1)$ representing that anisotropic parameter dominates, then $w_{\rm eff}=w_{\sigma}=1$, meaning the Universe behaves as if dominated by a stiff matter due to anisotropic expansion. The eigenvalues are $\{0, 0, -\sqrt{\frac{3}{2}}X, \frac{\sqrt{6}}{n}X\}$. The two eigenvalues of this point are zero, making this point non-hyperbolic. Since the non-zero eigenvalues depend on $X$, thus for point A, there are two cases:  $X\not=0$, there are always different signs, which makes this point a saddle point. In another case, $X=0$, all of the eigenvalues are zero, which makes this point center or neutral stability.

\subsection{Point B (Any, 0, 0, 1)}

The fixed line $S$ allows the range $0 \leq S \leq 1$ with the fixed point $X=Y=0$, meaning the scalar field has no kinetic and potential energy, $Z=1$ corresponds to $\lambda\to\infty$. Since $\Omega_\phi=0$, the scalar field does not contribute to the energy density of the Universe at this point. The deceleration parameter $q>0$ for all values $S$ represents no accelerated expansion. Anisotropy may play a role. Isotropic limit $(S=0)$, the vector field density parameter dominates at this point and the effective EoS is $w_{\rm eff}=w_A=1/3$, corresponding to radiation-dominated. Anisotropic limit $(S=1)$, the Universe is completely anisotropic, and the effective EoS is $w_{\rm eff}=w_{\sigma}=1$, which corresponds to a stiff matter-like behavior. Since all of the eigenvalues are zero parts, this point is fully non-hyperbolic, implying that point B is a center point or neutral stability.

\subsection{Point C (1, 0, 0, Any)}

This point represents a universe dominated by anisotropic expansion $(S=1)$. Since $X=Y=0$, the scalar field has no kinetic and potential energy. Both the energy density of the vector field and the scalar field are zero. The deceleration parameter at this point is $q=2$, implying that a high deceleration Universe. The effective equation of state corresponds directly to $w_\sigma$, which takes a value of $w_\sigma=1$, indicating a stiff matter behavior. Although the fixed line on the $Z$-axis is not dominated by any particular solution, it affects stability consideration. Since the non-zero eigenvalues depend on $Z$ that is bounded as $0 \leq Z \leq 1$. If $Z=0$ there are always opposite signs corresponding to saddle points. If $Z=1$ the stability is a center point.

\subsection{Point D ($\sqrt{1-X^2}$, Any, 0, 0)}

The fixed point $S=\sqrt{1-X^2}$ implies that the anisotropy depends on $X$, where $X$ is free to vary within $-1 \leq X \leq 1$, but $Y=0$ means that the scalar field does not have potential energy. Since $\Omega_A=0$ represents a universe without a vector field at this point. $\Omega_\phi$ depends on kinetic energy term only. In addition, this fixed point can be interpreted as a Kasner-like singularity \cite{Clarkson:2001fd}. The deceleration parameter at this point is $q=2$, which corresponds to a strongly decelerating Universe. The effective equation of state is $w_{\rm eff}=X^2+w_\sigma(1-X^2)$. If $X=0$, the universe is purely anisotropic with $w_{\rm eff}=w_\sigma=1$ (stiff matter due to anisotropic expansion). If $X\not=0$, the universe is purely dominated by the scalar field kinetic term, yielding $w_{\rm eff}=1$ (stiff matter). This point has two zero eigenvalues which are non-hyperbolic points. The non-zero eigenvalues are $\{2+6\sqrt{2}X-4\sqrt{1-X^2},3\}$, since they have at least one eigenvalue with a positive part, the corresponding fixed point can never be an attractor. If $X \leq 0$, two eigenvalues will always bear the opposite sign which makes these points a saddle point. If $X=1$ all of the non-zero eigenvalues will be positive parts which makes this point unstable (source).

\subsection{Point E (0, 0, 1, 0)}

The last fixed point E, represents the scalar field potential energy-dominated solution, characterized by an EoS $w_{\rm eff}=w_\phi=-1$. At this point $q<0$, the universe undergoes accelerated expansion. Since the fixed point E has one zero eigenvalue, rendering it non-hyperbolic, and all non-zero eigenvalues are negative, we employ the Center Manifold Theory (CMT) method to analyze its stability. To apply the CMT, we first shift this point $(0,0,1,0)$ to become the origin $(0,0,0,0)$ with the new set of variables given by
\begin{eqnarray}
    s &=& S\\
    x &=& X\\
    y &=& Y - 1\\
    z &=& Z
\end{eqnarray}
The autonomous system (\ref{autonomous sysytem1}-\ref{autonomous sysytem4}) with shifted this fixed point to the origin of manifold become:
\begin{eqnarray}
s' &=& (1-z) \left((s-2) \left(s^2+x^2-1\right)-2 (s+1) (y+1)^2\right)
\label{asE1}
\\
x' &=& (1-z) \left(\left(x+3 \sqrt{2}\right) \left(s^2+x^2+(y+1)^2-1\right)-3 x (y+1)^2+\frac{\sqrt{\frac{3}{2}} (y+1)^2z}{1-z}\right)
\label{asE2}
\\
y' &=& \frac{1}{2} (y+1) \left(-2 s^2 (z-1)-2 x^2 (z-1)-\sqrt{6} x z+4 \left((y+1)^2-1\right) (z-1)\right) 
\label{asE3}
\\
z' &=& \frac{\sqrt{6} x (z-1) z^2}{n}
\label{asE4}
\end{eqnarray}
The stability matrix evaluated at the origin is
\begin{eqnarray}
    J_E = \left(
\begin{array}{cccc}
 -3 & 0 & -4 & 0 \\
 0 & -3 & 6 \sqrt{2} & \sqrt{\frac{3}{2}} \\
 0 & 0 & -4 & 0 \\
 0 & 0 & 0 & 0 \\
\end{array}
\right)
\end{eqnarray}
The eigenvalues corresponding to this matrix are $\{-3,-3,-4,0\}$ with its eigenvectors $\left\{1,0,0,0\right\},\{0,1,0,0\},\\\{4,6\sqrt{2},1,0\}~\text{and}~\left\{0,\frac{1}{\sqrt{6}},0,1\right\}$. We can construct the transformation matrix from these eigenvectors, as 
\begin{eqnarray}
    T_E = \left(
\begin{array}{cccc}
 1 & 0 & 4 & 0 \\
 0 & 1 & -6 \sqrt{2} & \frac{1}{\sqrt{6}} \\
 0 & 0 & 1 & 0 \\
 0 & 0 & 0 & 1 \\
\end{array}
\right)
\end{eqnarray}
The new coordinate system, $(\mathrm{u},\mathrm{v},\mathrm{w},\mathrm{z})$ is defined by 
\begin{eqnarray}
\left(
\begin{array}{c}
 \mathrm{u} \\
 \mathrm{v} \\
 \mathrm{w} \\
 \mathrm{z} \\
\end{array}
\right) = T_E^{-1}\left(
\begin{array}{c}
 s \\
 x \\
 y \\
 z \\
\end{array}
\right)
\end{eqnarray}
We express the coordinate transformation as follows:
\begin{eqnarray}
\mathrm{u} &=& s-4y \\
\mathrm{v} &=& x+6\sqrt{2}y-{1\over\sqrt{6}}z\\
\mathrm{w} &=& y \\
\mathrm{z} &=& z
\end{eqnarray}
The autonomous system of the new variables can be represented in matrix form as
\begin{eqnarray}
    \left(
\begin{array}{c}
 \mathrm{u}' \\
 \mathrm{v}' \\
 \mathrm{w}' \\
 \mathrm{z}' \\
\end{array}
\right)=\left(
\begin{array}{cccc}
 -3 & 0 & 0 & 0 \\
 0 & -3 & 0 & 0 \\
 0 & 0 & -4 & 0 \\
 0 & 0 & 0 & 0 \\
\end{array}
\right)\left(
\begin{array}{c}
 \mathrm{u} \\
 \mathrm{v} \\
 \mathrm{w} \\
 \mathrm{z} \\
\end{array}
\right)+\left(
\begin{array}{c}
\text{non-linear terms}
\end{array}
\right)
\end{eqnarray}
Now, the coordinates corresponding to non-zero eigenvalues $(\mathrm{u},\mathrm{v},\mathrm{w})$ can be approximated in terms of the coordinate with a zero eigenvalue, such as $\mathrm{z}$ by the function $h(\mathrm{z})$ which given by
\begin{eqnarray}\label{h(z)forE}
    h(\mathrm{z})=
\left(
\begin{array}{c}
 \mathrm{u} \\
 \mathrm{v} \\
 \mathrm{w} \\
\end{array}
\right)=
\left(
\begin{array}{c}
 a_2\mathrm{z}^2+a_3\mathrm{z}^3+\mathcal{O}(\mathrm{z}^4) \\
 b_2\mathrm{z}^2+b_3\mathrm{z}^3+\mathcal{O}(\mathrm{z}^4) \\
 c_2\mathrm{z}^2+c_3\mathrm{z}^3+\mathcal{O}(\mathrm{z}^4) \\
\end{array}
\right)
\end{eqnarray}
By applying the quasilinear partial differential equation (\ref{tangencycond}) as
\begin{equation}
    \mathcal{N}(h(\mathrm{z})) \equiv Dh(\mathrm{z}) \left[ A z + f(\mathrm{z},h(\mathrm{z})) \right] - B h(\mathrm{z}) - g(\mathrm{z},h(\mathrm{z})) = 0
\end{equation}
where
\begin{equation}
    A = 0, \qquad
    B =
    \begin{pmatrix}
        -3 & 0 & 0 \\
        0 & -3 & 0 \\
        0 & 0 & -4
    \end{pmatrix}
\end{equation}
By comparing the coefficients of each power of $\mathrm{z}$, if the higher-order terms are very small, we can obtain:
\begin{eqnarray}\label{Econd}
    a_2 = \frac{1}{2}, \quad a_3 = 0, \quad
    b_2 = \frac{1}{6} \left( -3 \sqrt{2} + \sqrt{6} \right), \quad 
    b_3 = 0, \quad c_2 = -\frac{1}{12}, \quad c_3 = 0
    \label{compoCMT}
\end{eqnarray}
Substituting $h(\mathrm{z})$ into $\mathrm{z}$ and using (\ref{compoCMT}), we finally obtain the dynamics of the center manifold, given by:
\begin{eqnarray}
    \mathrm{z}'= -\frac{\mathrm{z}^3}{n} + \frac{\mathrm{z}^4}{n} - \frac{\sqrt{6}\,\mathrm{z}^4}{n} + \mathcal{O} (\mathrm{z^5})
\end{eqnarray}
This determines the stability of the reduced system, where the lowest power of $\mathrm{z}$ is an odd number corresponding to a stable fixed point for positive $n$. This analysis confirms that Point E is a stable attractor of the full system, driving the universe into an inflationary phase.

\subsection{Summary table: Fixed points and their stability}

From the analysis of each fixed point by evaluating the eigenvalues obtained from the Jacobian matrix, the stability of each fixed point can be summarized as shown in Table \ref{table:3}.

\begin{figure}[H]
    \centering    \includegraphics[width=0.6\linewidth]{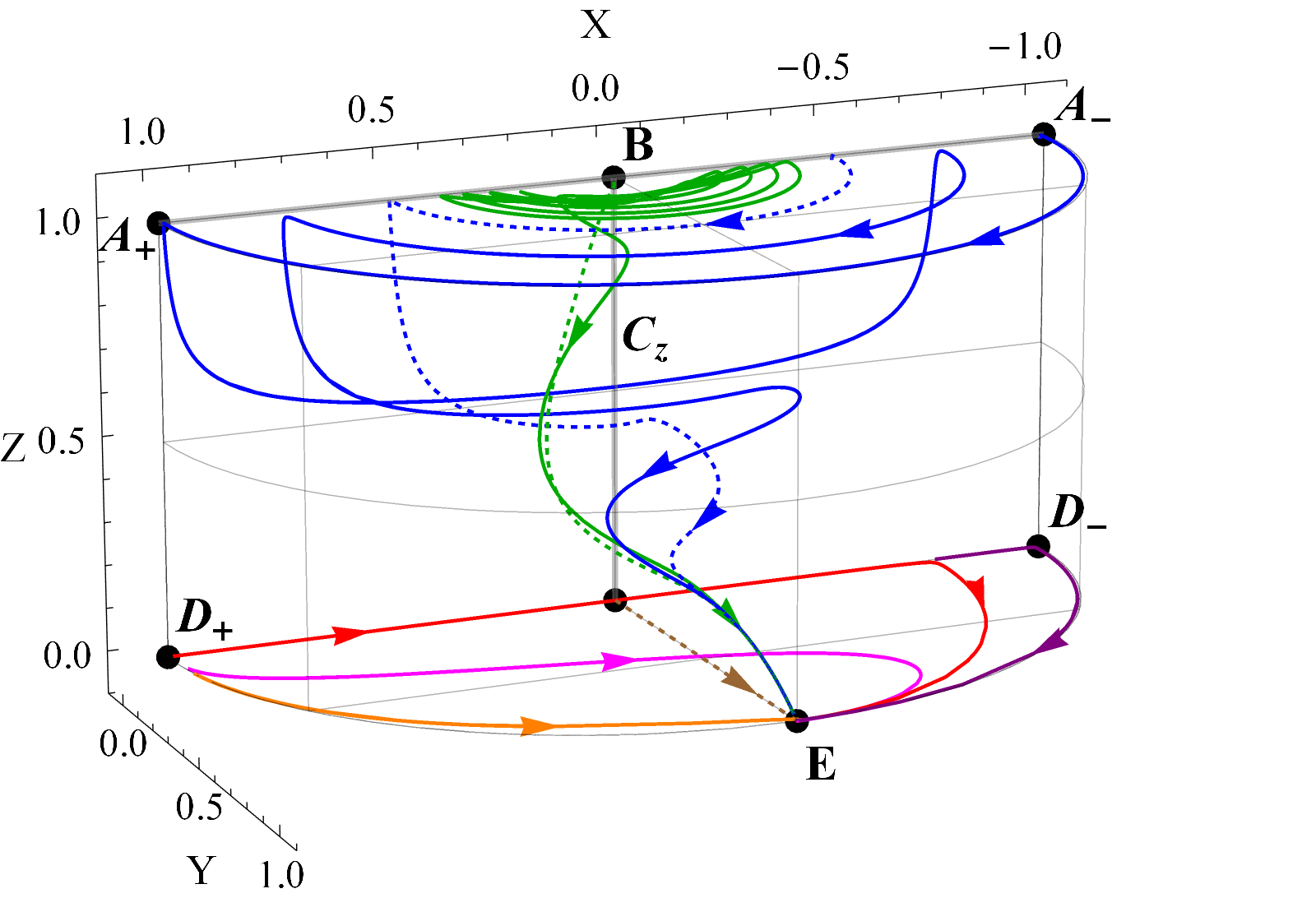}
    \caption{Phase space portrait of the system's dynamics in the $(X,Y,Z)$ coordinates. Trajectories (colored lines with arrows) show the evolution towards a stable attractor at point E potential scalar energy dominates. The different colors represent different initial conditions. Dashed lines represent trajectories starting from the anisotropic initial condition. The blue trajectory starts with initial conditions near the saddle point A$_{+/-}$, which is dominated by scalar kinetic energy. The green trajectory starts near the center point B. The purple trajectory begins close to the saddle point D$_-$, while the red, magenta, and orange trajectories start near the unstable point D$_+$, which is also dominated by scalar kinetic energy.}
    \label{fig:3Dtraject}
\end{figure}

\begin{table}[h!]
\setlength{\tabcolsep}{7pt}
\renewcommand{\arraystretch}{2.5}
\begin{tabular}{| c | c | c | c | c | c | c |}
\hline
Point & S & X & Y & Z & Eigenvalues & Stability\\
\hline

A & Any & Any & 0 & 1 & $\Big\{0,0,-{\sqrt{3\over2}}X,{\sqrt{6}\over n}X\Big\}$ & \makecell{Center for $X=0$ \\ Saddle for $X^2 \geq 1$}
\\

B & Any & 0 & 0 & 1 & $\{0,0,0,0\}$ & Center \\

C & 1 & 0 & 0 & Any & $\{2(Z-1),0,-3(Z-1),0\}$ & \makecell{Center for $Z=1$ \\ Saddle for $Z=0$}
\\

D & $\sqrt{1-X^2}$ & Any & 0 & 0 & $\{0,2+6\sqrt{2}X-4\sqrt{1-X^2},3,0)\}$ & \makecell{Unstable for $X=1$ \\ Saddle for $-1 \leq X \leq 0$}
\\

E & 0 & 0 & 1 & 0 & $\{-3,-3,-4,0\}$ & Stable\\
\hline
\end{tabular}
\caption{The stability analysis of fixed points using the Center Manifold Theorem.}
\label{table:3}
\end{table}
To demonstrate the evolution of a dynamical system, we numerically solve the autonomous equations with different initial conditions and then plot them to show the trajectories and evolution of the system. In FIG \ref{fig:3Dtraject} we plot the phase flow with different sets of specific initial conditions as different colored lines with parameter $n=4$. The solid lines represent the isotropy initial condition $(S=0)$, and the dashed lines are for the anisotropy property initial condition. From the solution, it can be seen that the unstable point is point D$_+$, which is dominated by the kinetic scalar. The universe cannot remain at this point for long. The fact that trajectories (red, magenta, and orange lines) escape from this point and eventually converge to point E indicates that even if the universe begins in a state of high kinetic energy, it can still be driven away from this point and undergo inflation. The saddle points are A$_{+/-}$, D$_-$, and C$_z$ for $Z=0$. The trajectories pass through these points on their way to point E, regardless of whether they begin with anisotropy. The green and blue trajectories (both solid and dashed lines) oscillate around the center point B, but they can eventually escape and move toward point E. This suggests that the center point B may act as a center only in the $X-Z$ plane, while in other directions, there are attractors or repulsive forces that cause the trajectories to move away from this point.

\begin{figure}[H]
    \centering    \includegraphics[width=0.6\linewidth]{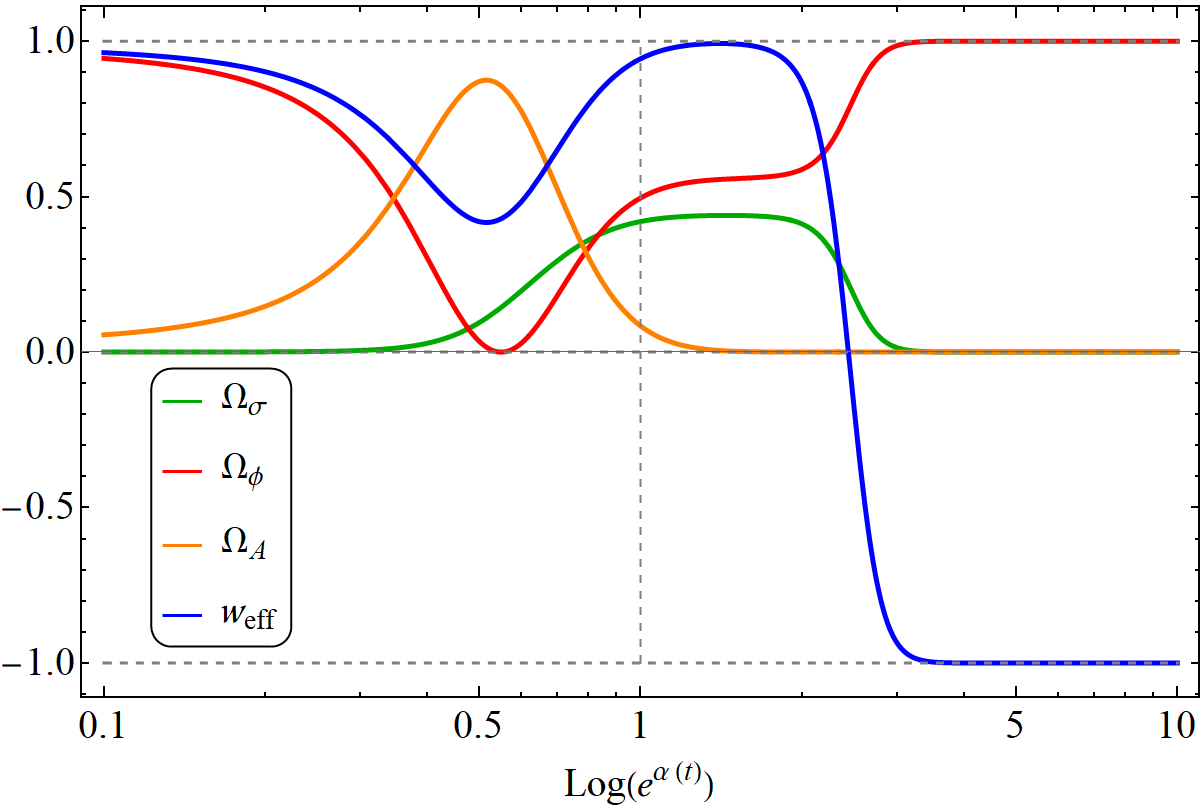}
    \caption{The evolution of the density parameters along with the scale factor. Initially, the universe begins with a scalar kinetic dominating (red). The vector field's energy density then grows (orange). Eventually, both the anisotropic (green) and vector field contributions decay, and the scalar field drives the universe into inflation, as indicated by effective EoS (blue) approaching $w_{\rm eff}\approx-1$.}
    \label{fig:evol}
\end{figure}

When the trajectory shown in red line is investigate, the evolution in FIG \ref{fig:evol} shows the evolution of the anisotropy contribution (green line), scalar field density parameter (red line), gauge field density parameter (orange line), and the effective equation of state (blue line) with respect to logarithm of the scale factor. This red line in FIG \ref{fig:3Dtraject} begins at the unstable point with a scalar kinetic dominated, then it passes through a region where the gauge field dominates, introducing anisotropy, although its contribution remains smaller than the scalar field energy density. This indicates that the universe originates from this point D$_+$, exhibiting a Kasner-like singularity due to the anisotropic dynamics driven by this unstable state. Moreover, this fixed point also appeared in Ref. \cite{Clarkson:2001fd} where they considered the magnetic Bianchi type I universe with exponential potential.  Eventually, the scalar field density quantity increases until the Universe is completely dominated by a scalar field with potential energy, undergoing inflation while becoming isotropic.

\section{Discussion and conclusion}
\textcolor{black}{
In this work, we have constructed and analyzed an anisotropic inflationary model inspired by KK theory, incorporating a gauge vector field and an inflaton (dilaton) field with the inverse power-law potential within the Bianchi type-I cosmological background. Our analysis employed the dynamical systems and CMT to investigate the stability of the model’s non-hyperbolic fixed points. The main findings of our study provide better understanding into the isotropization mechanism of the early universe and offer a theoretical motivation of the framework to explain how anisotropy could evolve into the homogeneity observed in the CMB.
\\
\indent
Our investigation reveals that Point E is an interesting fixed point where the scalar field potential dominates, 
while the vector and shear contributions vanish. This point corresponds to an isotropic de Sitter-like inflationary phase. However, this fixed point is the non-hyperbolic fixed point. By applying the CMT, we demonstrated that this fixed point is a true attractor in the phase space under broad initial conditions, including those with significant anisotropy or scalar kinetic energy dominance. More importantly, the stability of Point E is established through a cubic-order term in the center manifold dynamics, confirming its robustness under small perturbations.
\\
\indent
The numerical results support these analytical findings in this work. Trajectories in the phase space starting from both isotropic and highly anisotropic initial conditions were shown to converge toward Point E. Even trajectories that temporarily evolve through regions of vector field dominance or shear anisotropy are eventually driven toward the inflationary attractor. This result aligns well with the cosmic no-hair conjecture and affirms that the KK-inspired inverse power-law potential supports a dynamical isotropization mechanism. 
\\
\indent
Furthermore, the intermediate behavior of other fixed points particularly Points A, B, C, and D provides physical insight into the transient regimes that the universe may experience prior to entering the final inflationary phase. Point A and Point D, depending on the initial kinetic energy, exhibit stiff-matter-like behaviors with high deceleration parameters, acting as saddle or repelling regions in the phase space. Point B, a center point, allows for oscillatory evolution in certain directions but lacks attraction along the full dynamical trajectory. These features reflect the rich and varied pre-inflationary dynamics of anisotropic universes.
\\
\indent
From a theoretical perspective, our model highlights the natural emergence of a vector-inflaton system within the dimensional reduction of 5D KK gravity, reinforcing the physical plausibility of such models. The exponential coupling between the vector field and the dilaton emerges from the KK reduction procedure and plays a central role in the evolution of the anisotropy. The fact that such a structure yields a viable and stable isotropic inflationary solution provides strong motivation for further exploration of higher-dimensional and string-inspired models with non-trivial vector-scalar couplings.
\\
\indent
The use of the inverse power-law potential in this work is particularly interesting. Unlike exponential potentials commonly studied in the context of scaling solutions or assisted inflation, inverse power-law forms provide a graceful inflationary exit and are often associated with quintessential inflation and late-time acceleration scenarios. Our work demonstrates the feasibility of embedding such potentials in early-universe models that also address anisotropy and extra-dimensional physics.
\\
\indent
In conclusion, the KK-inspired vector-inflaton model with an inverse power-law potential in a Bianchi type-I universe provides a compelling and physically consistent scenario for the onset of inflation from anisotropic initial conditions. The combination of dynamical systems analysis and CMT confirms that isotropic inflation can emerge as a late-time attractor. These findings not only increase the general validity of the cosmic no-hair conjecture in extended gravity theories but also open avenues for further studies involving observational imprints of early-time anisotropies, such as potential statistical anisotropies in the CMB or polarization modes in primordial gravitational waves. Future work may include exploring the perturbation spectra, examining reheating dynamics in this framework, and extending the model to more general Bianchi types or higher-dimensional compactification schemes.}

\acknowledgments
DS is supported by the Fundamental Fund of Khon Kaen University and DS has received funding support from the National Science, Research and Innovation Fund. CP and DS are supported by Thailand NSRF via PMU-B [grant number B39G680009]. CP is also supported by Fundamental Fund 2568 of Khon Kaen University and Research Grant for New Scholar, Office of the Permanent Secretary, Ministry of Higher Education, Science, Research and Innovation under contract no. RGNS 64-043. PM, CP, and DS are partially supported by the National Astronomical Research Institute of Thailand (NARIT).

\bibliography{biblio}

\begin{thebibliography}{60}
\expandafter\ifx\csname natexlab\endcsname\relax\def\natexlab#1{#1}\fi
\expandafter\ifx\csname bibnamefont\endcsname\relax
  \def\bibnamefont#1{#1}\fi
\expandafter\ifx\csname bibfnamefont\endcsname\relax
  \def\bibfnamefont#1{#1}\fi
\expandafter\ifx\csname citenamefont\endcsname\relax
  \def\citenamefont#1{#1}\fi
\expandafter\ifx\csname url\endcsname\relax
  \def\url#1{\texttt{#1}}\fi
\expandafter\ifx\csname urlprefix\endcsname\relax\def\urlprefix{URL }\fi
\providecommand{\bibinfo}[2]{#2}
\providecommand{\eprint}[2][]{\url{#2}}

\bibitem[{\citenamefont{Clifton et~al.}(2012)\citenamefont{Clifton, Ferreira, Padilla, and Skordis}}]{Clifton:2011jh}
\bibinfo{author}{\bibfnamefont{T.}~\bibnamefont{Clifton}}, \bibinfo{author}{\bibfnamefont{P.~G.} \bibnamefont{Ferreira}}, \bibinfo{author}{\bibfnamefont{A.}~\bibnamefont{Padilla}}, \bibnamefont{and} \bibinfo{author}{\bibfnamefont{C.}~\bibnamefont{Skordis}}, \bibinfo{journal}{Phys. Rept.} \textbf{\bibinfo{volume}{513}}, \bibinfo{pages}{1} (\bibinfo{year}{2012}), \eprint{1106.2476}.

\bibitem[{\citenamefont{Gron}(1985)}]{Gron:1985hw}
\bibinfo{author}{\bibfnamefont{O.}~\bibnamefont{Gron}}, \bibinfo{journal}{Phys. Rev. D} \textbf{\bibinfo{volume}{32}}, \bibinfo{pages}{1586} (\bibinfo{year}{1985}), \bibinfo{note}{[Erratum: Phys.Rev.D 34, 664 (1986)]}.

\bibitem[{\citenamefont{Lim et~al.}(2004)\citenamefont{Lim, van Elst, Uggla, and Wainwright}}]{Lim:2003ut}
\bibinfo{author}{\bibfnamefont{W.~C.} \bibnamefont{Lim}}, \bibinfo{author}{\bibfnamefont{H.}~\bibnamefont{van Elst}}, \bibinfo{author}{\bibfnamefont{C.}~\bibnamefont{Uggla}}, \bibnamefont{and} \bibinfo{author}{\bibfnamefont{J.}~\bibnamefont{Wainwright}}, \bibinfo{journal}{Phys. Rev. D} \textbf{\bibinfo{volume}{69}}, \bibinfo{pages}{103507} (\bibinfo{year}{2004}), \eprint{gr-qc/0306118}.

\bibitem[{\citenamefont{Pereira and Pitrou}(2015)}]{Pereira:2015pga}
\bibinfo{author}{\bibfnamefont{T.}~\bibnamefont{Pereira}} \bibnamefont{and} \bibinfo{author}{\bibfnamefont{C.}~\bibnamefont{Pitrou}}, \bibinfo{journal}{Comptes Rendus Physique} \textbf{\bibinfo{volume}{16}}, \bibinfo{pages}{1027} (\bibinfo{year}{2015}), \eprint{1509.09166}.

\bibitem[{\citenamefont{Nojiri et~al.}(2022)\citenamefont{Nojiri, Odintsov, Oikonomou, and Constantini}}]{Nojiri:2022idp}
\bibinfo{author}{\bibfnamefont{S.}~\bibnamefont{Nojiri}}, \bibinfo{author}{\bibfnamefont{S.~D.} \bibnamefont{Odintsov}}, \bibinfo{author}{\bibfnamefont{V.~K.} \bibnamefont{Oikonomou}}, \bibnamefont{and} \bibinfo{author}{\bibfnamefont{A.}~\bibnamefont{Constantini}}, \bibinfo{journal}{Nucl. Phys. B} \textbf{\bibinfo{volume}{985}}, \bibinfo{pages}{116011} (\bibinfo{year}{2022}), \eprint{2210.16383}.

\bibitem[{\citenamefont{Ford}(1989)}]{PhysRevD.40.967}
\bibinfo{author}{\bibfnamefont{L.~H.} \bibnamefont{Ford}}, \bibinfo{journal}{Phys. Rev. D} \textbf{\bibinfo{volume}{40}}, \bibinfo{pages}{967} (\bibinfo{year}{1989}), \urlprefix\url{https://link.aps.org/doi/10.1103/PhysRevD.40.967}.

\bibitem[{\citenamefont{Golovnev et~al.}(2008)\citenamefont{Golovnev, Mukhanov, and Vanchurin}}]{Golovnev:2008cf}
\bibinfo{author}{\bibfnamefont{A.}~\bibnamefont{Golovnev}}, \bibinfo{author}{\bibfnamefont{V.}~\bibnamefont{Mukhanov}}, \bibnamefont{and} \bibinfo{author}{\bibfnamefont{V.}~\bibnamefont{Vanchurin}}, \bibinfo{journal}{JCAP} \textbf{\bibinfo{volume}{06}}, \bibinfo{pages}{009} (\bibinfo{year}{2008}), \eprint{0802.2068}.

\bibitem[{\citenamefont{Kanno et~al.}(2008)\citenamefont{Kanno, Kimura, Soda, and Yokoyama}}]{Kanno:2008gn}
\bibinfo{author}{\bibfnamefont{S.}~\bibnamefont{Kanno}}, \bibinfo{author}{\bibfnamefont{M.}~\bibnamefont{Kimura}}, \bibinfo{author}{\bibfnamefont{J.}~\bibnamefont{Soda}}, \bibnamefont{and} \bibinfo{author}{\bibfnamefont{S.}~\bibnamefont{Yokoyama}}, \bibinfo{journal}{JCAP} \textbf{\bibinfo{volume}{08}}, \bibinfo{pages}{034} (\bibinfo{year}{2008}), \eprint{0806.2422}.

\bibitem[{\citenamefont{Watanabe et~al.}(2009)\citenamefont{Watanabe, Kanno, and Soda}}]{Watanabe:2009ct}
\bibinfo{author}{\bibfnamefont{M.-a.} \bibnamefont{Watanabe}}, \bibinfo{author}{\bibfnamefont{S.}~\bibnamefont{Kanno}}, \bibnamefont{and} \bibinfo{author}{\bibfnamefont{J.}~\bibnamefont{Soda}}, \bibinfo{journal}{Phys. Rev. Lett.} \textbf{\bibinfo{volume}{102}}, \bibinfo{pages}{191302} (\bibinfo{year}{2009}), \eprint{0902.2833}.

\bibitem[{\citenamefont{Kanno et~al.}(2010)\citenamefont{Kanno, Soda, and Watanabe}}]{Kanno:2010nr}
\bibinfo{author}{\bibfnamefont{S.}~\bibnamefont{Kanno}}, \bibinfo{author}{\bibfnamefont{J.}~\bibnamefont{Soda}}, \bibnamefont{and} \bibinfo{author}{\bibfnamefont{M.-a.} \bibnamefont{Watanabe}}, \bibinfo{journal}{JCAP} \textbf{\bibinfo{volume}{12}}, \bibinfo{pages}{024} (\bibinfo{year}{2010}), \eprint{1010.5307}.

\bibitem[{\citenamefont{Soda}(2012)}]{Soda:2012zm}
\bibinfo{author}{\bibfnamefont{J.}~\bibnamefont{Soda}}, \bibinfo{journal}{Class. Quant. Grav.} \textbf{\bibinfo{volume}{29}}, \bibinfo{pages}{083001} (\bibinfo{year}{2012}), \eprint{1201.6434}.

\bibitem[{\citenamefont{Ito and Soda}(2018)}]{Ito:2017bnn}
\bibinfo{author}{\bibfnamefont{A.}~\bibnamefont{Ito}} \bibnamefont{and} \bibinfo{author}{\bibfnamefont{J.}~\bibnamefont{Soda}}, \bibinfo{journal}{Eur. Phys. J. C} \textbf{\bibinfo{volume}{78}}, \bibinfo{pages}{55} (\bibinfo{year}{2018}), \eprint{1710.09701}.

\bibitem[{\citenamefont{Chen and Soda}(2021)}]{Chen:2021nkf}
\bibinfo{author}{\bibfnamefont{C.-B.} \bibnamefont{Chen}} \bibnamefont{and} \bibinfo{author}{\bibfnamefont{J.}~\bibnamefont{Soda}}, \bibinfo{journal}{JCAP} \textbf{\bibinfo{volume}{09}}, \bibinfo{pages}{026} (\bibinfo{year}{2021}), \eprint{2106.04813}.

\bibitem[{\citenamefont{Murata and Soda}(2011)}]{Murata:2011wv}
\bibinfo{author}{\bibfnamefont{K.}~\bibnamefont{Murata}} \bibnamefont{and} \bibinfo{author}{\bibfnamefont{J.}~\bibnamefont{Soda}}, \bibinfo{journal}{JCAP} \textbf{\bibinfo{volume}{06}}, \bibinfo{pages}{037} (\bibinfo{year}{2011}), \eprint{1103.6164}.

\bibitem[{\citenamefont{Maleknejad and Sheikh-Jabbari}(2013)}]{Maleknejad:2011jw}
\bibinfo{author}{\bibfnamefont{A.}~\bibnamefont{Maleknejad}} \bibnamefont{and} \bibinfo{author}{\bibfnamefont{M.~M.} \bibnamefont{Sheikh-Jabbari}}, \bibinfo{journal}{Phys. Lett. B} \textbf{\bibinfo{volume}{723}}, \bibinfo{pages}{224} (\bibinfo{year}{2013}), \eprint{1102.1513}.

\bibitem[{\citenamefont{Maleknejad et~al.}(2013)\citenamefont{Maleknejad, Sheikh-Jabbari, and Soda}}]{Maleknejad:2012fw}
\bibinfo{author}{\bibfnamefont{A.}~\bibnamefont{Maleknejad}}, \bibinfo{author}{\bibfnamefont{M.~M.} \bibnamefont{Sheikh-Jabbari}}, \bibnamefont{and} \bibinfo{author}{\bibfnamefont{J.}~\bibnamefont{Soda}}, \bibinfo{journal}{Phys. Rept.} \textbf{\bibinfo{volume}{528}}, \bibinfo{pages}{161} (\bibinfo{year}{2013}), \eprint{1212.2921}.

\bibitem[{\citenamefont{Ito and Soda}(2015)}]{Ito:2015sxj}
\bibinfo{author}{\bibfnamefont{A.}~\bibnamefont{Ito}} \bibnamefont{and} \bibinfo{author}{\bibfnamefont{J.}~\bibnamefont{Soda}}, \bibinfo{journal}{Phys. Rev. D} \textbf{\bibinfo{volume}{92}}, \bibinfo{pages}{123533} (\bibinfo{year}{2015}), \eprint{1506.02450}.

\bibitem[{\citenamefont{Goodarzi}(2022)}]{Goodarzi:2022wli}
\bibinfo{author}{\bibfnamefont{P.}~\bibnamefont{Goodarzi}}, \bibinfo{journal}{JCAP} \textbf{\bibinfo{volume}{11}}, \bibinfo{pages}{052} (\bibinfo{year}{2022}), \eprint{2208.10757}.

\bibitem[{\citenamefont{Kanno et~al.}(2023)\citenamefont{Kanno, Mukuno, Soda, and Ueda}}]{Kanno:2022flo}
\bibinfo{author}{\bibfnamefont{S.}~\bibnamefont{Kanno}}, \bibinfo{author}{\bibfnamefont{A.}~\bibnamefont{Mukuno}}, \bibinfo{author}{\bibfnamefont{J.}~\bibnamefont{Soda}}, \bibnamefont{and} \bibinfo{author}{\bibfnamefont{K.}~\bibnamefont{Ueda}}, \bibinfo{journal}{Phys. Rev. D} \textbf{\bibinfo{volume}{107}}, \bibinfo{pages}{063524} (\bibinfo{year}{2023}), \eprint{2209.05776}.

\bibitem[{\citenamefont{Pham et~al.}(2024)\citenamefont{Pham, Nguyen, Do, and Kao}}]{Pham:2023evo}
\bibinfo{author}{\bibfnamefont{T.~M.} \bibnamefont{Pham}}, \bibinfo{author}{\bibfnamefont{D.~H.} \bibnamefont{Nguyen}}, \bibinfo{author}{\bibfnamefont{T.~Q.} \bibnamefont{Do}}, \bibnamefont{and} \bibinfo{author}{\bibfnamefont{W.~F.} \bibnamefont{Kao}}, \bibinfo{journal}{Eur. Phys. J. C} \textbf{\bibinfo{volume}{84}}, \bibinfo{pages}{105} (\bibinfo{year}{2024}), \eprint{2309.02690}.

\bibitem[{\citenamefont{Pitrou et~al.}(2008)\citenamefont{Pitrou, Pereira, and Uzan}}]{Pitrou:2008gk}
\bibinfo{author}{\bibfnamefont{C.}~\bibnamefont{Pitrou}}, \bibinfo{author}{\bibfnamefont{T.~S.} \bibnamefont{Pereira}}, \bibnamefont{and} \bibinfo{author}{\bibfnamefont{J.-P.} \bibnamefont{Uzan}}, \bibinfo{journal}{JCAP} \textbf{\bibinfo{volume}{04}}, \bibinfo{pages}{004} (\bibinfo{year}{2008}), \eprint{0801.3596}.

\bibitem[{\citenamefont{Soda}(2015)}]{Soda:2014awa}
\bibinfo{author}{\bibfnamefont{J.}~\bibnamefont{Soda}}, \bibinfo{journal}{J. Phys. Conf. Ser.} \textbf{\bibinfo{volume}{600}}, \bibinfo{pages}{012026} (\bibinfo{year}{2015}), \eprint{1410.8643}.

\bibitem[{\citenamefont{Do and Kao}(2021)}]{Do:2021lyf}
\bibinfo{author}{\bibfnamefont{T.~Q.} \bibnamefont{Do}} \bibnamefont{and} \bibinfo{author}{\bibfnamefont{W.~F.} \bibnamefont{Kao}}, \bibinfo{journal}{Eur. Phys. J. C} \textbf{\bibinfo{volume}{81}}, \bibinfo{pages}{525} (\bibinfo{year}{2021}), \eprint{2104.14100}.

\bibitem[{\citenamefont{Do and Kao}(2022)}]{Do:2021pqk}
\bibinfo{author}{\bibfnamefont{T.~Q.} \bibnamefont{Do}} \bibnamefont{and} \bibinfo{author}{\bibfnamefont{W.~F.} \bibnamefont{Kao}}, \bibinfo{journal}{Eur. Phys. J. C} \textbf{\bibinfo{volume}{82}}, \bibinfo{pages}{123} (\bibinfo{year}{2022}), \eprint{2110.13516}.

\bibitem[{\citenamefont{Do and Kao}(2024)}]{Do:2023mqe}
\bibinfo{author}{\bibfnamefont{T.~Q.} \bibnamefont{Do}} \bibnamefont{and} \bibinfo{author}{\bibfnamefont{W.~F.} \bibnamefont{Kao}}, \bibinfo{journal}{Phys. Scripta} \textbf{\bibinfo{volume}{99}}, \bibinfo{pages}{015002} (\bibinfo{year}{2024}), \eprint{2304.08874}.

\bibitem[{\citenamefont{Do et~al.}(2025)\citenamefont{Do, Dong, Nguyen, and Singh}}]{Do:2025xhm}
\bibinfo{author}{\bibfnamefont{T.~Q.} \bibnamefont{Do}}, \bibinfo{author}{\bibfnamefont{P.~V.} \bibnamefont{Dong}}, \bibinfo{author}{\bibfnamefont{D.~H.} \bibnamefont{Nguyen}}, \bibnamefont{and} \bibinfo{author}{\bibfnamefont{J.~K.} \bibnamefont{Singh}} (\bibinfo{year}{2025}), \eprint{2502.10462}.

\bibitem[{\citenamefont{Overduin and Wesson}(1997)}]{Overduin:1997sri}
\bibinfo{author}{\bibfnamefont{J.~M.} \bibnamefont{Overduin}} \bibnamefont{and} \bibinfo{author}{\bibfnamefont{P.~S.} \bibnamefont{Wesson}}, \bibinfo{journal}{Phys. Rept.} \textbf{\bibinfo{volume}{283}}, \bibinfo{pages}{303} (\bibinfo{year}{1997}), \eprint{gr-qc/9805018}.

\bibitem[{\citenamefont{Bailin and Love}(1987)}]{Bailin:1987jd}
\bibinfo{author}{\bibfnamefont{D.}~\bibnamefont{Bailin}} \bibnamefont{and} \bibinfo{author}{\bibfnamefont{A.}~\bibnamefont{Love}}, \bibinfo{journal}{Rept. Prog. Phys.} \textbf{\bibinfo{volume}{50}}, \bibinfo{pages}{1087} (\bibinfo{year}{1987}).

\bibitem[{\citenamefont{Freund}(1982)}]{Freund:1982pg}
\bibinfo{author}{\bibfnamefont{P.~G.~O.} \bibnamefont{Freund}}, \bibinfo{journal}{Nucl. Phys. B} \textbf{\bibinfo{volume}{209}}, \bibinfo{pages}{146} (\bibinfo{year}{1982}).

\bibitem[{\citenamefont{Randjbar-Daemi et~al.}(1984)\citenamefont{Randjbar-Daemi, Salam, and Strathdee}}]{Randjbar-Daemi:1983awk}
\bibinfo{author}{\bibfnamefont{S.}~\bibnamefont{Randjbar-Daemi}}, \bibinfo{author}{\bibfnamefont{A.}~\bibnamefont{Salam}}, \bibnamefont{and} \bibinfo{author}{\bibfnamefont{J.~A.} \bibnamefont{Strathdee}}, \bibinfo{journal}{Phys. Lett. B} \textbf{\bibinfo{volume}{135}}, \bibinfo{pages}{388} (\bibinfo{year}{1984}).

\bibitem[{\citenamefont{Osorio and Vazquez-Mozo}(1993)}]{Osorio:1993iv}
\bibinfo{author}{\bibfnamefont{M.~A.~R.} \bibnamefont{Osorio}} \bibnamefont{and} \bibinfo{author}{\bibfnamefont{M.~A.} \bibnamefont{Vazquez-Mozo}}, \bibinfo{journal}{Mod. Phys. Lett. A} \textbf{\bibinfo{volume}{8}}, \bibinfo{pages}{3111} (\bibinfo{year}{1993}), \eprint{hep-th/9305137}.

\bibitem[{\citenamefont{Sahoo and Mishra}(2014)}]{Sahoo:2014wgz}
\bibinfo{author}{\bibfnamefont{P.~K.} \bibnamefont{Sahoo}} \bibnamefont{and} \bibinfo{author}{\bibfnamefont{B.}~\bibnamefont{Mishra}}, \bibinfo{journal}{Can. J. Phys.} \textbf{\bibinfo{volume}{92}}, \bibinfo{pages}{1062} (\bibinfo{year}{2014}).

\bibitem[{\citenamefont{Pongkitivanichkul et~al.}(2020)\citenamefont{Pongkitivanichkul, Samart, Thongyoi, and Lunrasri}}]{Pongkitivanichkul:2020txi}
\bibinfo{author}{\bibfnamefont{C.}~\bibnamefont{Pongkitivanichkul}}, \bibinfo{author}{\bibfnamefont{D.}~\bibnamefont{Samart}}, \bibinfo{author}{\bibfnamefont{N.}~\bibnamefont{Thongyoi}}, \bibnamefont{and} \bibinfo{author}{\bibfnamefont{N.}~\bibnamefont{Lunrasri}}, \bibinfo{journal}{Phys. Dark Univ.} \textbf{\bibinfo{volume}{30}}, \bibinfo{pages}{100731} (\bibinfo{year}{2020}), \eprint{2005.08791}.

\bibitem[{\citenamefont{Surendra~Singh et~al.}(2021)\citenamefont{Surendra~Singh, Manihar~Singh, and Kumrah}}]{SurendraSingh:2021uxb}
\bibinfo{author}{\bibfnamefont{S.}~\bibnamefont{Surendra~Singh}}, \bibinfo{author}{\bibfnamefont{K.}~\bibnamefont{Manihar~Singh}}, \bibnamefont{and} \bibinfo{author}{\bibfnamefont{L.}~\bibnamefont{Kumrah}}, \bibinfo{journal}{Int. J. Mod. Phys. A} \textbf{\bibinfo{volume}{36}}, \bibinfo{pages}{2150043} (\bibinfo{year}{2021}).

\bibitem[{\citenamefont{Waeming et~al.}(2022)\citenamefont{Waeming, Klangburam, Pongkitivanichkul, and Samart}}]{Waeming:2021ytf}
\bibinfo{author}{\bibfnamefont{A.}~\bibnamefont{Waeming}}, \bibinfo{author}{\bibfnamefont{T.}~\bibnamefont{Klangburam}}, \bibinfo{author}{\bibfnamefont{C.}~\bibnamefont{Pongkitivanichkul}}, \bibnamefont{and} \bibinfo{author}{\bibfnamefont{D.}~\bibnamefont{Samart}}, \bibinfo{journal}{Eur. Phys. J. C} \textbf{\bibinfo{volume}{82}}, \bibinfo{pages}{409} (\bibinfo{year}{2022}), \eprint{2107.00678}.

\bibitem[{\citenamefont{Jusufi et~al.}(2025)\citenamefont{Jusufi, Luciano, Sheykhi, and Samart}}]{Jusufi:2024utf}
\bibinfo{author}{\bibfnamefont{K.}~\bibnamefont{Jusufi}}, \bibinfo{author}{\bibfnamefont{G.~G.} \bibnamefont{Luciano}}, \bibinfo{author}{\bibfnamefont{A.}~\bibnamefont{Sheykhi}}, \bibnamefont{and} \bibinfo{author}{\bibfnamefont{D.}~\bibnamefont{Samart}}, \bibinfo{journal}{JHEAp} \textbf{\bibinfo{volume}{47}}, \bibinfo{pages}{100373} (\bibinfo{year}{2025}), \eprint{2411.14176}.

\bibitem[{\citenamefont{Garc\'\i{}a-Salcedo et~al.}(2015)\citenamefont{Garc\'\i{}a-Salcedo, Gonzalez, Horta-Rangel, Quiros, and Sanchez-Guzm\'an}}]{Garcia-Salcedo:2015ora}
\bibinfo{author}{\bibfnamefont{R.}~\bibnamefont{Garc\'\i{}a-Salcedo}}, \bibinfo{author}{\bibfnamefont{T.}~\bibnamefont{Gonzalez}}, \bibinfo{author}{\bibfnamefont{F.~A.} \bibnamefont{Horta-Rangel}}, \bibinfo{author}{\bibfnamefont{I.}~\bibnamefont{Quiros}}, \bibnamefont{and} \bibinfo{author}{\bibfnamefont{D.}~\bibnamefont{Sanchez-Guzm\'an}}, \bibinfo{journal}{Eur. J. Phys.} \textbf{\bibinfo{volume}{36}}, \bibinfo{pages}{025008} (\bibinfo{year}{2015}), \eprint{1501.04851}.

\bibitem[{\citenamefont{Tamanini}(2014)}]{Tamanini:2014mpa}
\bibinfo{author}{\bibfnamefont{N.}~\bibnamefont{Tamanini}}, \bibinfo{journal}{Phys. Rev. D} \textbf{\bibinfo{volume}{89}}, \bibinfo{pages}{083521} (\bibinfo{year}{2014}), \eprint{1401.6339}.

\bibitem[{\citenamefont{Alho et~al.}(2015)\citenamefont{Alho, Hell, and Uggla}}]{Alho:2015cza}
\bibinfo{author}{\bibfnamefont{A.}~\bibnamefont{Alho}}, \bibinfo{author}{\bibfnamefont{J.}~\bibnamefont{Hell}}, \bibnamefont{and} \bibinfo{author}{\bibfnamefont{C.}~\bibnamefont{Uggla}}, \bibinfo{journal}{Class. Quant. Grav.} \textbf{\bibinfo{volume}{32}}, \bibinfo{pages}{145005} (\bibinfo{year}{2015}), \eprint{1503.06994}.

\bibitem[{\citenamefont{Bouhmadi-L\'opez et~al.}(2017)\citenamefont{Bouhmadi-L\'opez, Marto, Morais, and Silva}}]{Bouhmadi-Lopez:2016dzw}
\bibinfo{author}{\bibfnamefont{M.}~\bibnamefont{Bouhmadi-L\'opez}}, \bibinfo{author}{\bibfnamefont{J.~a.} \bibnamefont{Marto}}, \bibinfo{author}{\bibfnamefont{J.~a.} \bibnamefont{Morais}}, \bibnamefont{and} \bibinfo{author}{\bibfnamefont{C.~M.} \bibnamefont{Silva}}, \bibinfo{journal}{JCAP} \textbf{\bibinfo{volume}{03}}, \bibinfo{pages}{042} (\bibinfo{year}{2017}), \eprint{1611.03100}.

\bibitem[{\citenamefont{Boehmer and Chan}(2017)}]{Boehmer:2014vea}
\bibinfo{author}{\bibfnamefont{C.~G.} \bibnamefont{Boehmer}} \bibnamefont{and} \bibinfo{author}{\bibfnamefont{N.}~\bibnamefont{Chan}}, \emph{\bibinfo{title}{{Dynamical systems in cosmology.}}} (\bibinfo{year}{2017}), \eprint{1409.5585}.

\bibitem[{\citenamefont{Bahamonde et~al.}(2018)\citenamefont{Bahamonde, B\"ohmer, Carloni, Copeland, Fang, and Tamanini}}]{Bahamonde:2017ize}
\bibinfo{author}{\bibfnamefont{S.}~\bibnamefont{Bahamonde}}, \bibinfo{author}{\bibfnamefont{C.~G.} \bibnamefont{B\"ohmer}}, \bibinfo{author}{\bibfnamefont{S.}~\bibnamefont{Carloni}}, \bibinfo{author}{\bibfnamefont{E.~J.} \bibnamefont{Copeland}}, \bibinfo{author}{\bibfnamefont{W.}~\bibnamefont{Fang}}, \bibnamefont{and} \bibinfo{author}{\bibfnamefont{N.}~\bibnamefont{Tamanini}}, \bibinfo{journal}{Phys. Rept.} \textbf{\bibinfo{volume}{775-777}}, \bibinfo{pages}{1} (\bibinfo{year}{2018}), \eprint{1712.03107}.

\bibitem[{\citenamefont{Pal}(2022)}]{Pal:2022jak}
\bibinfo{author}{\bibfnamefont{S.}~\bibnamefont{Pal}}, Ph.D. thesis, \bibinfo{school}{Jadavpur University, Department of Mathematics, India} (\bibinfo{year}{2022}).

\bibitem[{\citenamefont{Alho et~al.}(2022)\citenamefont{Alho, Lim, and Uggla}}]{Alho:2022ehq}
\bibinfo{author}{\bibfnamefont{A.}~\bibnamefont{Alho}}, \bibinfo{author}{\bibfnamefont{W.~C.} \bibnamefont{Lim}}, \bibnamefont{and} \bibinfo{author}{\bibfnamefont{C.}~\bibnamefont{Uggla}}, \bibinfo{journal}{Class. Quant. Grav.} \textbf{\bibinfo{volume}{39}}, \bibinfo{pages}{145010} (\bibinfo{year}{2022}), \eprint{2201.07596}.

\bibitem[{\citenamefont{Clarkson et~al.}(2001)\citenamefont{Clarkson, Coley, and Quinlan}}]{Clarkson:2001fd}
\bibinfo{author}{\bibfnamefont{C.~A.} \bibnamefont{Clarkson}}, \bibinfo{author}{\bibfnamefont{A.~A.} \bibnamefont{Coley}}, \bibnamefont{and} \bibinfo{author}{\bibfnamefont{S.~D.} \bibnamefont{Quinlan}}, \bibinfo{journal}{Phys. Rev. D} \textbf{\bibinfo{volume}{64}}, \bibinfo{pages}{122003} (\bibinfo{year}{2001}), \eprint{astro-ph/0108268}.

\bibitem[{\citenamefont{Copeland et~al.}(1998)\citenamefont{Copeland, Liddle, and Wands}}]{Copeland:1997et}
\bibinfo{author}{\bibfnamefont{E.~J.} \bibnamefont{Copeland}}, \bibinfo{author}{\bibfnamefont{A.~R.} \bibnamefont{Liddle}}, \bibnamefont{and} \bibinfo{author}{\bibfnamefont{D.}~\bibnamefont{Wands}}, \bibinfo{journal}{Phys. Rev. D} \textbf{\bibinfo{volume}{57}}, \bibinfo{pages}{4686} (\bibinfo{year}{1998}), \eprint{gr-qc/9711068}.

\bibitem[{\citenamefont{Heard and Wands}(2002)}]{Heard:2002dr}
\bibinfo{author}{\bibfnamefont{I.~P.~C.} \bibnamefont{Heard}} \bibnamefont{and} \bibinfo{author}{\bibfnamefont{D.}~\bibnamefont{Wands}}, \bibinfo{journal}{Class. Quant. Grav.} \textbf{\bibinfo{volume}{19}}, \bibinfo{pages}{5435} (\bibinfo{year}{2002}), \eprint{gr-qc/0206085}.

\bibitem[{\citenamefont{Guo et~al.}(2003)\citenamefont{Guo, Piao, and Zhang}}]{Guo:2003eu}
\bibinfo{author}{\bibfnamefont{Z.~K.} \bibnamefont{Guo}}, \bibinfo{author}{\bibfnamefont{Y.-S.} \bibnamefont{Piao}}, \bibnamefont{and} \bibinfo{author}{\bibfnamefont{Y.-Z.} \bibnamefont{Zhang}}, \bibinfo{journal}{Phys. Lett. B} \textbf{\bibinfo{volume}{568}}, \bibinfo{pages}{1} (\bibinfo{year}{2003}), \eprint{hep-th/0304048}.

\bibitem[{\citenamefont{Das et~al.}(2019)\citenamefont{Das, Banerjee, and Roy}}]{Das:2019ixt}
\bibinfo{author}{\bibfnamefont{S.}~\bibnamefont{Das}}, \bibinfo{author}{\bibfnamefont{M.}~\bibnamefont{Banerjee}}, \bibnamefont{and} \bibinfo{author}{\bibfnamefont{N.}~\bibnamefont{Roy}}, \bibinfo{journal}{JCAP} \textbf{\bibinfo{volume}{08}}, \bibinfo{pages}{024} (\bibinfo{year}{2019}), \eprint{1903.02288}.

\bibitem[{\citenamefont{Chaubey and Raushan}(2016)}]{Chaubey:2016qtx}
\bibinfo{author}{\bibfnamefont{R.}~\bibnamefont{Chaubey}} \bibnamefont{and} \bibinfo{author}{\bibfnamefont{R.}~\bibnamefont{Raushan}}, \bibinfo{journal}{Astrophys. Space Sci.} \textbf{\bibinfo{volume}{361}}, \bibinfo{pages}{215} (\bibinfo{year}{2016}).

\bibitem[{\citenamefont{Kar\v{c}iauskas}(2016)}]{Karciauskas:2016pxn}
\bibinfo{author}{\bibfnamefont{M.}~\bibnamefont{Kar\v{c}iauskas}}, \bibinfo{journal}{Mod. Phys. Lett. A} \textbf{\bibinfo{volume}{31}}, \bibinfo{pages}{1640002} (\bibinfo{year}{2016}), \eprint{1604.00269}.

\bibitem[{\citenamefont{Ghaffarnejad and Yaraie}(2017)}]{Ghaffarnejad:2017hqt}
\bibinfo{author}{\bibfnamefont{H.}~\bibnamefont{Ghaffarnejad}} \bibnamefont{and} \bibinfo{author}{\bibfnamefont{E.}~\bibnamefont{Yaraie}}, \bibinfo{journal}{Gen. Rel. Grav.} \textbf{\bibinfo{volume}{49}}, \bibinfo{pages}{49} (\bibinfo{year}{2017}), \eprint{1604.06269}.

\bibitem[{\citenamefont{Guarnizo et~al.}(2020)\citenamefont{Guarnizo, Orjuela-Quintana, and Valenzuela-Toledo}}]{Guarnizo:2020pkj}
\bibinfo{author}{\bibfnamefont{A.}~\bibnamefont{Guarnizo}}, \bibinfo{author}{\bibfnamefont{J.~B.} \bibnamefont{Orjuela-Quintana}}, \bibnamefont{and} \bibinfo{author}{\bibfnamefont{C.~A.} \bibnamefont{Valenzuela-Toledo}}, \bibinfo{journal}{Phys. Rev. D} \textbf{\bibinfo{volume}{102}}, \bibinfo{pages}{083507} (\bibinfo{year}{2020}), \eprint{2007.12964}.

\bibitem[{\citenamefont{Bhanja et~al.}(2023)\citenamefont{Bhanja, Mandal, Mamon, and Biswas}}]{Bhanja:2023jro}
\bibinfo{author}{\bibfnamefont{S.}~\bibnamefont{Bhanja}}, \bibinfo{author}{\bibfnamefont{G.}~\bibnamefont{Mandal}}, \bibinfo{author}{\bibfnamefont{A.~A.} \bibnamefont{Mamon}}, \bibnamefont{and} \bibinfo{author}{\bibfnamefont{S.~K.} \bibnamefont{Biswas}}, \bibinfo{journal}{JCAP} \textbf{\bibinfo{volume}{10}}, \bibinfo{pages}{050} (\bibinfo{year}{2023}), \eprint{2307.13000}.

\bibitem[{\citenamefont{Do et~al.}(2023)\citenamefont{Do, Nguyen, and Pham}}]{Do:2023yvg}
\bibinfo{author}{\bibfnamefont{T.~Q.} \bibnamefont{Do}}, \bibinfo{author}{\bibfnamefont{D.~H.} \bibnamefont{Nguyen}}, \bibnamefont{and} \bibinfo{author}{\bibfnamefont{T.~M.} \bibnamefont{Pham}}, \bibinfo{journal}{Int. J. Mod. Phys. D} \textbf{\bibinfo{volume}{32}}, \bibinfo{pages}{2350087} (\bibinfo{year}{2023}), \eprint{2303.17283}.

\bibitem[{\citenamefont{Rendall}(2002)}]{Rendall:2001it}
\bibinfo{author}{\bibfnamefont{A.~D.} \bibnamefont{Rendall}}, \bibinfo{journal}{Gen. Rel. Grav.} \textbf{\bibinfo{volume}{34}}, \bibinfo{pages}{1277} (\bibinfo{year}{2002}), \eprint{gr-qc/0112040}.

\bibitem[{\citenamefont{Alho and Uggla}(2015)}]{Alho:2014fha}
\bibinfo{author}{\bibfnamefont{A.}~\bibnamefont{Alho}} \bibnamefont{and} \bibinfo{author}{\bibfnamefont{C.}~\bibnamefont{Uggla}}, \bibinfo{journal}{J. Math. Phys.} \textbf{\bibinfo{volume}{56}}, \bibinfo{pages}{012502} (\bibinfo{year}{2015}), \eprint{1406.0438}.

\bibitem[{\citenamefont{Hrycyna and Szyd\l{}owski}(2015)}]{Hrycyna:2015eta}
\bibinfo{author}{\bibfnamefont{O.}~\bibnamefont{Hrycyna}} \bibnamefont{and} \bibinfo{author}{\bibfnamefont{M.}~\bibnamefont{Szyd\l{}owski}}, \bibinfo{journal}{JCAP} \textbf{\bibinfo{volume}{11}}, \bibinfo{pages}{013} (\bibinfo{year}{2015}), \eprint{1506.03429}.

\bibitem[{\citenamefont{Boehmer et~al.}(2012)\citenamefont{Boehmer, Chan, and Lazkoz}}]{Boehmer:2011tp}
\bibinfo{author}{\bibfnamefont{C.~G.} \bibnamefont{Boehmer}}, \bibinfo{author}{\bibfnamefont{N.}~\bibnamefont{Chan}}, \bibnamefont{and} \bibinfo{author}{\bibfnamefont{R.}~\bibnamefont{Lazkoz}}, \bibinfo{journal}{Phys. Lett. B} \textbf{\bibinfo{volume}{714}}, \bibinfo{pages}{11} (\bibinfo{year}{2012}), \eprint{1111.6247}.

\bibitem[{\citenamefont{Pal and Chakraborty}(2019)}]{Pal:2019qch}
\bibinfo{author}{\bibfnamefont{S.}~\bibnamefont{Pal}} \bibnamefont{and} \bibinfo{author}{\bibfnamefont{S.}~\bibnamefont{Chakraborty}}, \bibinfo{journal}{Gen. Rel. Grav.} \textbf{\bibinfo{volume}{51}}, \bibinfo{pages}{124} (\bibinfo{year}{2019}), \eprint{2103.02715}.

\end{thebibliography}

\end{document}